\documentclass{aa}

\usepackage{amsmath}
\usepackage{graphicx}
\usepackage{psfig}
\usepackage{txfonts}
\usepackage{textcomp}
\usepackage{natbib}

\def\arcsec{\hbox{$^{\prime\prime}$}}
\def\degr{\hbox{$^{\circ}$}}

\newcommand{\etal}{et~al. }

\newcommand{\caiih}{Ca\,{\sc ii}\,{\sc H}}

\newcommand{\SP}{{Solar Phys.}}

\newcommand{\Ha}{${\rm H\alpha}$ }
\newcommand{\ha}{${\rm H\alpha}$ }

\begin{document}

\title{Wave propagation in a solar quiet region and the influence of the magnetic canopy}

       \author{I.Kontogiannis\inst{}
        \and G.\,Tsiropoula\inst{}
        \and K.\,Tziotziou\inst{}}

\institute{Institute for Astronomy, Astrophysics, Space Applications
and Remote Sensing, National Observatory of Athens,
GR-15236 Penteli, Greece\\
        \email{[jkonto; georgia; kostas]@noa.gr}}

\offprints{I.\,Kontogiannis\\ \email{jkonto@noa.gr}}

\date{Received <date> / Accepted <accepted>}

\abstract {}
{We seek indications or evidence of transmission/conversion
of magnetoacoustic waves at the magnetic canopy, as a result of its impact on the
properties of the wave field of the photosphere and chromosphere.}
{We use cross-wavelet analysis to measure phase differences between intensity and Doppler signal oscillations in the H$\alpha$, \caiih,\ and G-band. We use the height of the magnetic canopy to
create appropriate masks to separate internetwork (IN)\ and magnetic canopy regions.
We study wave propagation and differences between these two regions.}
{The magnetic canopy affects wave propagation by lowering
the phase differences of progressive waves and allowing the
propagation of waves with frequencies lower than the acoustic
cut-off. We also find indications in the Doppler signals of H$\alpha$ of
a response to the acoustic waves at the IN, observed in the \caiih\ line. This response is affected by the presence of the magnetic canopy.}
{Phase difference analysis indicates the existence of a
complicated wave field in the quiet Sun, which is composed of a mixture of
progressive and standing waves. There are clear
imprints of mode conversion and transmission due to the
interaction between the p-modes and small-scale magnetic fields
of the network and internetwork.}

\titlerunning{phase differences in quiet Sun}
\authorrunning{Kontogiannis \etal}
\maketitle

\section{Introduction}
Wave propagation has been a long standing field of research in solar
physics, spanning several decades of literature. The discovery of
the ``5-min oscillations'' by \citet{leighton} and the observation
and study of the p-mode spectrum using ``k-$\omega$'' diagrams by
\citet{deub75} and \citet{deub79} showed that there is a pool of
acoustic oscillations, with periods of a few minutes that permeate
the solar interior and atmosphere. These acoustic oscillations are capable of transporting
and depositing energy and may also give valuable information
on local physical conditions. Several aspects concerning wave
propagation are established through a series of papers
\citep[e.g.][]{mein76,lites79,lites82,kneer85,fleck89,deub89,deub_fleck90,deub90,lites93,krijger}.
One of the tools used for diagnosing the details of wave propagation is
phase difference analysis, which facilitates the comparison of velocity and intensity signals
that form at the same or different atmospheric layers. In the traditional view of wave propagation in the solar atmosphere,
waves with frequencies lower than 5.2\,mHz cannot propagate from the
photosphere to the chromosphere. This theoretical property has been
verified by the small phase differences measured between oscillations at two heights.
Currently, owing to high-resolution observations, this description is being reassessed to
include wave propagation with respect to the local magnetic field topology, since
the latter is playing a crucial role in wave processes at the solar
atmosphere.

In contrast to its definition, quiet Sun continuously and
ubiquitously exhibits a large range of dynamic phenomena associated with small scale, albeit intense,
magnetic field concentrations. Pushed by convective and near surface motions, magnetic fields accumulate at the
boundaries of supergranules and form the magnetic network, visible
as a bright web-like pattern (in filtergrams of strong chromospheric
lines such as \caiih) surrounding larger areas of very low magnetic
flux called internetwork (IN). These bright points betray, in fact,
the presence of magnetic flux tubes, which expand with height, as
the ambient atmospheric pressure drops, and result in highly
inclined magnetic fields that dominate the dynamics of the overlying
chromosphere. Observationally, in the \ha line (6563\,{\AA}) this
magnetic configuration manifests itself through the ubiquitous
presence of elongated absorption features, called mottles
\citep{beckers68,tsirop12}, which stem from the network and partially cover
the IN. Mottles form the so-called magnetic canopy
\citep{gab76,kont10b}, which roughly separates the stratified,
gas-pressure dominated environment of the photosphere and lower
chromosphere and the
plasma of the upper atmosphere, which is highly structured by the magnetic field. The plasma-$\beta$, denoting the
ratio of the gas pressure to the magnetic pressure, is used to
parameterize the plasma dynamics, while the height at which
$\beta=1$ defines the height of the magnetic canopy.

\begin{figure*}[ht]
\centering
\includegraphics[width=12 cm]{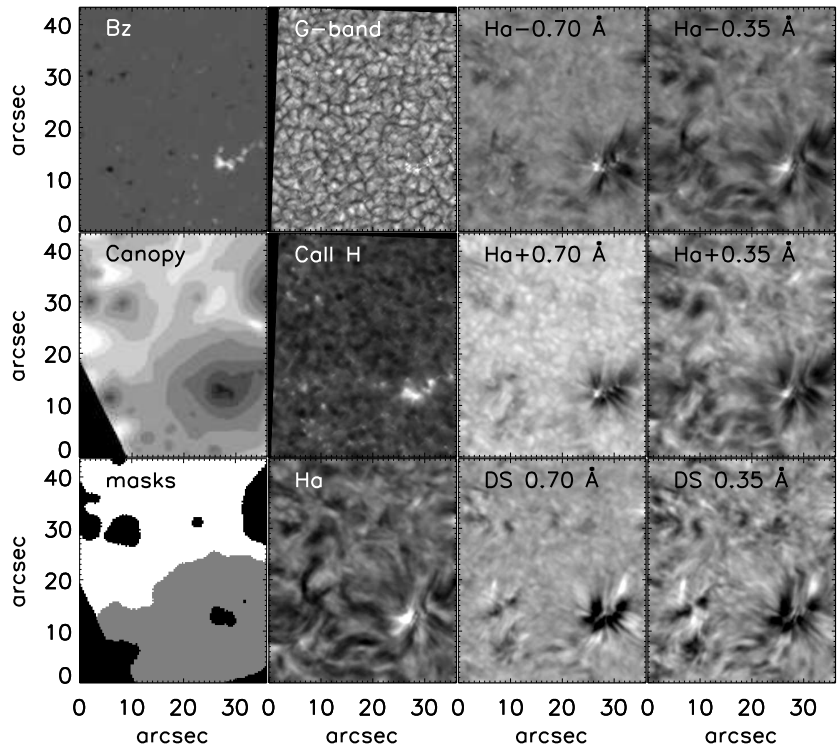}
\caption{Summary of the observations of the quiet Sun area under
study. In the left panel, top row, white denotes positive
polarity. In the left panel, middle row, lighter shading
denotes higher canopy while canopy heights start at 250\,km and
increase in 235\,km steps. In the left panel, bottom row,
white denotes the mask of the IN area, grey the mask of the canopy
area while the black regions have not been taken under
consideration. Images and DS maps are temporal averages. At the DS
maps of the bottom row, white denotes upwards and black
denotes downwards motion.} \label{data}
\end{figure*}

The interaction of the acoustic oscillations with the magnetic field
of the network and the magnetic canopy has been illustrated in a
series of studies. In general, the magnetic field may allow or
inhibit the propagation of waves of certain frequencies. The presence
of both intense and inclined magnetic field may lower the acoustic
cut-off frequency, allowing the propagation of  waves with frequencies
lower than 5.2\,mHz \citep{mich73,bel,suematsu}; this may explain the detection of
5\,min oscillations in chromospheric mottles, which is a common
finding \citep{tziotziou04,tsirop09}. The chromospheric magnetic field is also responsible for the so-called ``power halos'' and ``magnetic shadows'', areas of increased and
suppressed power, respectively, found over and around the network
\citep{krijger,judge,mcintosh01,mcintosh03,vecchio,reardon09,kont10a}.
This was also clearly demonstrated in \citet{kont10b,kontogiannis14}
who concluded that waves reflect and refract on the magnetic canopy,
increasing the power below and decreasing it above.

Details of the mechanism that produces wave reflection and refraction on the magnetic canopy
have been highlighted in several theoretical works and simulations with increasing level of
sophistication \citep{ros02,bog03,carlsson06,khom06,khom08,khom09,nutto}. Following
these studies, it is clear that the critical parameter for the
interaction between the acoustic waves and the magnetic canopy is
the attack angle, defined by the inclination of the magnetic field
vector and the direction of the wave-vector. Upon hitting the
magnetic canopy, the acoustic waves transfer part of their energy to
a slow magnetoacoustic wave (a process termed transmission) and part
to a fast magnetoacoustic wave (a process termed conversion). The amount of energy
that goes to each of these modes has been defined in the model of
\citet{schunker06} and \citet{cally07} and depends on the attack
angle, the frequency of the wave, and the thickness of the magnetic
canopy. Application on solar observations explains very well the
distribution of acoustic power around sunspots \citep{stang11} and
the network \citep{kontogiannis14}.

As a consequence of this kind of interaction, waves in the
chromosphere are strongly guided by the magnetic field. Slow waves
travel along the magnetic field lines, while fast waves propagate
perpendicularly and eventually reflect, propagating downwards and forming standing waves. This sequence of events probably results to a complicated wave field,
since standing and propagating waves (both vertically and obliquely)
are superposed to a different extent, depending on the height and
position in the solar atmosphere. In this paper we measure the phase differences of oscillations using time series of filtergrams that represent
different atmospheric heights and dynamic regimes. Our aim is to
examine whether the differences between the IN and the magnetic canopy, if any, are consistent with the scheme of transmission and conversion of magneto-acoustic waves on the magnetic canopy.

\section{Observations}

Our data were obtained on October 15, 2007, as part of an
observational campaign, which included several ground-based and
space-born instruments. We used data from the Dutch
Open Telescope \citep[DOT;][]{rutten2004a} and the Spectropolarimeter of the Solar Optical Telescope \citep[SOT/SP;][]{sot} on board \textit{Hinode} \citep{hinode}.

The DOT observed a 84\arcsec$\times$87\arcsec area located at the
centre of the solar disk between 08:32--09:53 UT and provided a time
series of speckle reconstructed images in five positions along the
\ha line profile (line centre, $\pm$0.35\,{\AA}, $\pm$0.70\,{\AA}),
as well as \caiih\ and G-band filtergrams. The cadence of the time
series is 30\,s and the spatial resolution is 0.109\,arcsec/pixel.
The spectral coverage of the \ha profile is implemented throughout
the 30\,s run and, therefore, images at different wavelengths of the
profile have been acquired with a small time difference. To
compensate for this effect, we spline-interpolated all filtergrams to
a common time axis and in the following we consider that they
were all acquired simultaneously.

The intensities at $\pm0.35\,{\AA}$ and $\pm0.70\,{\AA}$ from the \ha line centre, are used to calculate the corresponding Doppler signals (DS), through the formula
  \begin{equation}
    \label{eq:1}
 { DS=\frac{I(+\Delta\lambda)-I(-\Delta\lambda)}{I(+\Delta\lambda)+I(-\Delta\lambda)}\;\;\;.    }
  \end{equation}
\noindent The DS gives a parametric description of the velocity with
upwards motion denoted by positive values and vice versa. Also, the
averaged wing intensity may be calculated by averaging the intensity at
opposite wings, i.e.
  \begin{equation}
    \label{eq:2}
 { I_{avg}(\Delta\lambda)=\frac{I(+\Delta\lambda)+I(-\Delta\lambda)}{2}.    }
  \end{equation}
\noindent This gives a more reliable intensity measure at the
corresponding wing position, compensated for the effect of Doppler
shift.

The SOT/SP performed a raster scan of the same region in the Fe\,I
6301.5 and 6302.5\,{\AA} lines (in normal mode), with a spatial
resolution of 0.32\,arcsec/pixel between 09:05--09:15 UT. The
corresponding vector magnetogram was produced
by the Community Spectropolarimetric Analysis Center of the High Altitude Observatory (HAO/CSAC), through inversion of the Stokes spectra via the
MERLIN code. Further details on the observations, the speckle
reconstruction procedure of the \ha observations, data reduction
steps, and the inversion of the Stokes spectra can be found in
\citet{rutten2004a} and \citet{kont10b}.

Standard data reduction procedures were used, with a considerable
effort devoted to image co-alignment, through cross-correlation, between
average images at different filters (Fig.~\ref{data}). In the common field of view (FOV) of 36\arcsec$\times$44\arcsec\ we  studied, a well-defined rosette of mottles is found around a network boundary cluster of positive polarity magnetic fields (Fig.~\ref{data}). Since we are planning to study wave propagation inside certain areas of the FOV, a few limitations arise: the examined time series must have no gaps in image acquisition and must be co-temporal with high-resolution magnetograms. The part of the time series that satisfies both conditions is between 09:00--09:30\,UT.

\section{Analysis}

\subsection{Determination of the height of the magnetic canopy}
The SOT/SP raster was used as the lower photospheric boundary condition
to calculate the chromospheric magnetic field through a current-free (potential field) extrapolation (Schmidt, 1964) up to 2500\,km. The potential extrapolation gives
the vector magnetic field at the chromosphere, which corresponds to
the minimum current-free energy state \citep[for details
see][]{kont10b,kontogiannis11}. Then the plasma-$\beta$
parameter ($\beta=P_{gas}/P_{mag}$) is calculated, where $P_{gas}$, the gas pressure, is taken from model C of \citet{vernazza} and $P_{mag}$ is the
magnetic pressure ($B^{2}/2\mu_{0}$). The height of the
magnetic canopy was then determined, as the height where
$\beta$ is of order unity \citep[see][]{kont10b}.

To study the effect of the magnetic canopy on wave propagation, the FOV was divided in two parts, using the height of the magnetic canopy as a criterion. We chose this classification based on the comparison between the height of the magnetic canopy and the power maps of 3 and 5\,min acoustic oscillations from  \citet{kont10b}.
We consider as ``IN'' the part of the FOV where the magnetic canopy is
higher than 1600\,km and as ``canopy'', the area around the
well-formed rosette, where the height of the magnetic canopy is between 720\,km and 1600\,km. We have excluded from our analysis the network boundaries, where the magnetic canopy
forms lower than 720\,km as well as some low-lying canopies found at the IN
(Fig.~\ref{data}).

\subsection{Phase difference analysis}

We examine wave propagation at the IN and the magnetic canopy
through a cross-wavelet transform \citep{torr} between intensities,
DS, or intensity-DS pairs on every pixel of the FOV. Although phase differences may be obtained with a fast Fourier transform (widely utilized in past studies),  wavelet analysis was chosen as a more suitable method because of the intermittency exhibited by the solar oscillations. We used the Morlet wavelet function, which is a sinusoid modulated by a Gaussian and is appropriate and commonly used for harmonic-type oscillations.

While a large part of the literature has been based on the ``k-$\omega$'' diagrams, including studies on the global nature of solar oscillations persisting up to chromospheric heights \citep{kneer85,deub96}, for reasons of consistency with our previous works and to facilitate comparison between results, our analysis is restricted only to frequencies. This approach has been widely used in nominal works referring to network-IN oscillations \citep[e.g.][]{mein76,lites79,lites82,deub90,lites93}. All of these studies examined the vertical propagation of acoustic waves. This, of course, does not mean that acoustic waves propagate only vertically. In the last section, we will discuss the limitations posed by this assumption and the kind of analysis and data sets needed to overcome these limitations.  Thus, in the present analysis we focus on the interaction between the vertically propagating acoustic waves with the inclined magnetic field lines that form the magnetic canopy in an attempt to provide further supporting evidence to our findings regarding the role of the magnetic canopy in the transmission/conversion of magnetoacoustic waves at the network/IN areas \citep{kontogiannis14}.

A cross-wavelet analysis of two signals
gives the crosspower, coherence, and phase difference
as functions of time and period (frequency). Coherence is
a measure of the cross-correlation between the two time series and
assumes values from zero (where no correlation exists between the
two time series) to unity. Random noise, however, produces a
non-zero coherence and therefore, a coherence threshold must be
determined, above which the phase difference is considered reliable.
This task is not trivial in the case of the wavelet transform.
In order to remedy this, we adopt the ``floor exceedance coherency approach''
described in \citet{bloomfield04}. The resulting coherence threshold
for all time series is frequency dependent and, in most cases, it is around 0.6.

We discarded all phase differences that correspond to lower coherence values. For the
remaining phase differences, we follow the methodology of \citet{lites79} by constructing halftone images, where the crosspower at every pixel of the area is summed over
3$\degr$ wide bins of phase difference for each frequency element (calculated for the periods of the wavelet transform). To improve the readability of these images, the crosspower distribution for each frequency is normalized to unity.

We should stress that through this methodology the phase differences themselves are not averaged inside the corresponding areas. Instead, the resulting 2-D graphs represent the distribution of the crosspower measured in each pixel of the canopy/IN areas,  frequency, and phase difference bins, and these 2-D graphs serve two purposes. First, they exhibit the trend of phase difference as a function of frequency. To further enhance this feature, we also calculated the position of the maximum crosspower for each frequency and overlaid its position on each distribution. Second, the width of the distribution gives a measure of the scatter of crosspower \citep{lites93}. To provide a measure of this scatter, we overlay the 50\% contour of the halftone images as an estimate for the phase difference distribution width (representing the FWHM of the crosspower distribution at each frequency). The difference in our approach, compared to previous studies, is that we utilize all available crosspower inside the canopy/IN areas to construct the halftone images, provided that it corresponds to coherence higher than the previously derived threshold.

\begin{figure*}[htp]
\centering
\includegraphics[width=19 cm]{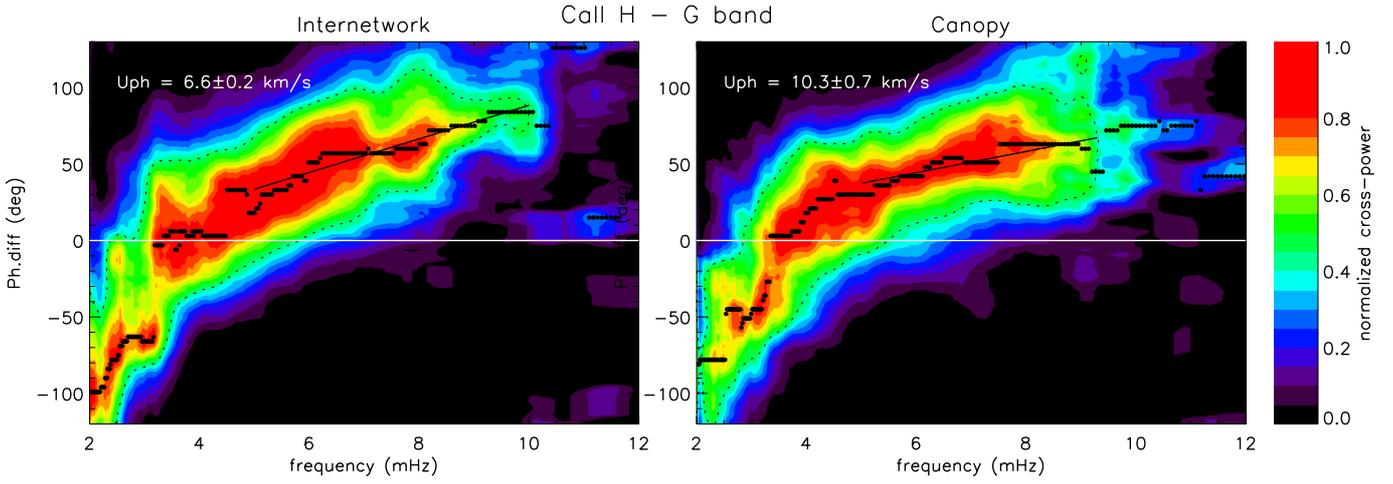}
\caption[]{Phase differences between \caiih\--G-band at the IN (left) and the magnetic canopy (right). Filled contours represent the crosspower distribution, black points denote
the positions of maximum crosspower, normalized to unity for each
frequency element, and the dashed contour indicates the 50\% level from
maximum crosspower (as an indicator of FWHM). Crosspower below 10\%
has been disregarded. Also, overplotted are regression lines calculated above $5.2\,mHz$ up to the part of the distribution enclosed by the 0.5 contour. From their gradient, the corresponding phase velocities are calculated, assuming that the height separation between \caiih\ and G-band is
200\,km (see text).}
\label{ca_gb}
\end{figure*}

For vertical wave propagation between two atmospheric
levels and for frequencies higher than the acoustic cut-off, the
dependence of phase difference on frequency is given by:
  \begin{equation}
    \label{eq:4}
 { \Delta\phi=2\pi\frac{dh}{\upsilon_{ph}}f  }
  ,\end{equation}
\noindent where $f$ is the frequency, $\upsilon_{ph}$ the phase
speed, and $dh$ the height separation between the height of formation
(HOF) of the considered intensity/velocity pair. Therefore, when measuring
phase differences between two fixed heights, a monotonic increase of phase differences
is an indication of upwards vertical propagation. For a given height
separation, the steeper the gradient, the lower the phase speed, and
one may use the gradient to determine the average phase speed. On the other hand,
small and constant
phase differences are an indication of standing or evanescent waves. The
interpretation of phase differences between intensity and velocity
signals within the same spectral line is more complicated because of the
necessary simplifications adopted concerning the spectral line
formation details and the dependence of brightness on the
thermodynamic parameters of the emitting/absorbing gas (i.e. temperature,density). Several theoretical studies have addressed this issue in the past and we
attempt to utilize their results.

\section{Results}

\subsection{Phase differences between \caiih\ and G-band}

First, we examine wave propagation between the photosphere in the G-band and the layers sampled by the \caiih\ bandpass. This  also gives us the opportunity to test our approach since this pair has been studied before \citep[see e.g.][]{rutten2004b,lawrence12}. The wide bandpass of the \caiih\ filter contains contribution from the core of the line, which is of chromospheric origin. This contribution, however, is overshadowed by the photospheric part of the line \citep{reardon09} and, therefore, the bandpass mostly shows  the reversed granulation and the network boundary. The DOT \caiih\ intensity is estimated by \citet{rutten2004b} to reflect heights around 250\,km. On the  other hand, the G-band is formed up to a few tens of km above $\tau_{5000}=1,$ so it is reasonable to assume a height separation between the two layers of around 200\,km.

Figure~\ref{ca_gb} shows the phase differences between the two bandpasses at the IN and the canopy. In line with previous studies, both at the IN and the canopy regions, the negative phase differences at the low-frequency range (below 3\,mHz) are attributed to gravity waves \citep{mihalas81} and the intensity anti-correlation between \caiih\ and G-band due to the reversed granulation.

At the IN (Fig.\ref{ca_gb}, left panel), phase differences between 3\,mHz and 5\,mHz are constant and around 0$\degr$, indicating the presence of evanescent waves. Then, they start to increase at $\sim$4.5\,mHz, while the width of the distribution shows significant cross--power corresponding to positive phase difference at frequencies as low as 3\,mHz. The inference of a cut-off frequency lower than 5.2\,mHz at the IN is an interesting finding, suggesting the presence of small-scale, unresolved inclined magnetic fields. This is in agreement with the latest observations that reveal magnetic fields at the IN, stronger than previously thought \citep{lites08}. Indeed, an inspection of the original SOT/SP raster magnetogram shows a multitude of small-scale LOS magnetic elements at the IN with measured strengths around 100\,G. The presence of strong magnetic fields in very small scales cannot be ruled out because these fields would be smeared by the resolution element, since magnetic flux, instead of the field strength, is measured. However, even in the absence of strong magnetic fields, a lower cut-off value is possible as the result of radiative damping in an isothermal atmosphere \citep[see e.g.][]{worrall02}. Above 5\,mHz, phase differences increase monotonically with frequency. A linear fit on the maxima of the distribution leads to an estimation of the phase speed at 6.6$\pm$0.2\,$kms^{-1}$, which is compatible with the sound speed at the photosphere.

\begin{figure*}[ht]
\centering
\includegraphics[width=19 cm]{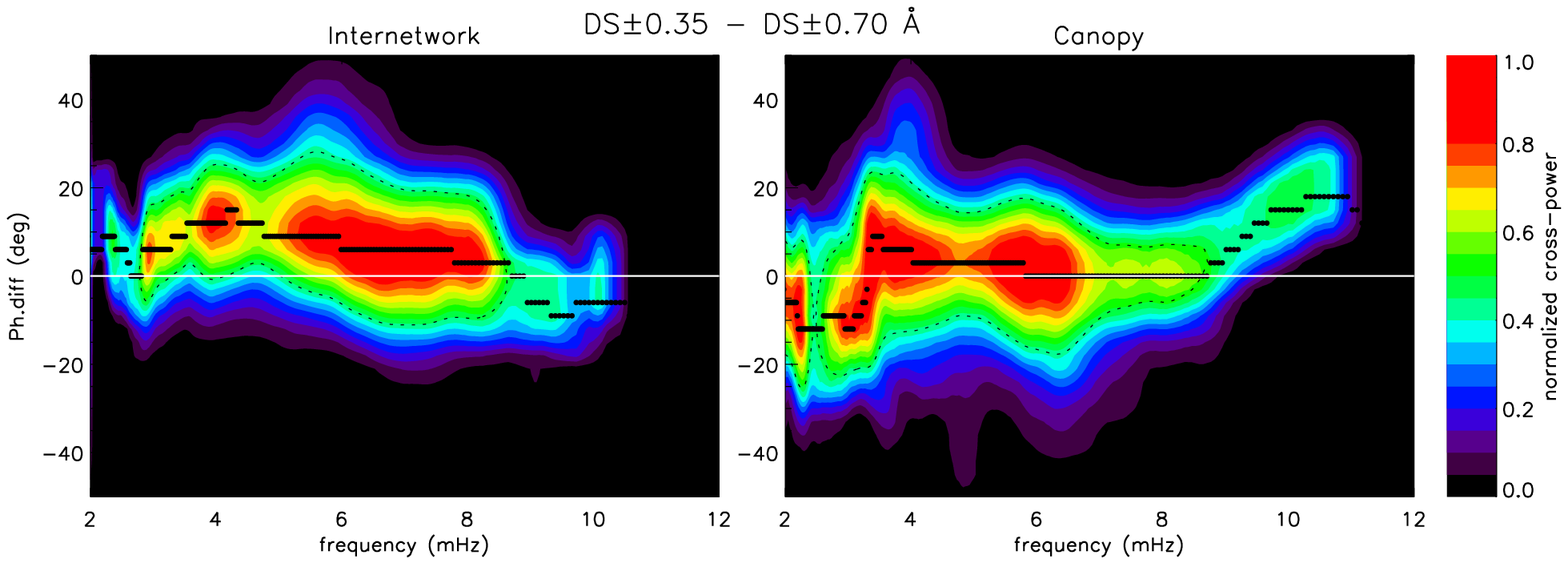}
\caption[]{Same as Fig.~\ref{ca_gb}, for the pair \ha DS at $\pm$0.35\,{\AA} and
$\pm$0.70\,{\AA}.}
\label{ds1_ds2}
\end{figure*}

At the magnetic canopy (Fig.\ref{ca_gb}, right panel), phase differences start increasing below 4\,mHz. This is an indication of the lowering of the acoustic cut-off frequency in the vicinity of the network boundaries \citep{mich73}. Phase differences increase with frequency, above 4\,mHz, as expected for acoustic wave propagation. The gradient of the distribution is lower than that at the IN and leads to a higher phase speed (10.3$\pm$0.7\,$kms^{-1}$). A higher phase speed in the vicinity of the magnetic canopy is explained on the grounds of mode conversion according to which part of the energy carried by the acoustic waves is transferred to the fast magneto-acoustic mode. The fast speed increases near the network boundary, as the strength of the magnetic field increases there as well.

Lower phase differences near the network boundary may also be produced by the decrease of the height separation between the two layers as a result of the presence of the diverging magnetic field. This scenario was invoked by \citet{mcintosh04} to explain the dependence of wave packet characteristics on the magnetic field strength. \citet{nutto}, using realistic MHD simulations of the quiet Sun, conclude that indeed it is the combination of this effect with the increasing phase speed of the fast mode due to the presence of the magnetic canopy, which produces shortened travel times, i.e. lower phase differences. The suppression of acoustic power around magnetic concentrations in \caiih\ observations \citep{lawrence2010} shows that the magnetic canopy affects the oscillatory field in this bandpass and, therefore, the latter mechanism may also be in effect as well.

The effect of the magnetic canopy in our data is not so pronounced as in the case of \citet{lawrence12}. The reason lies in our mask selection and the fact that in our FOV, there is a highly localized network boundary instead of an extended distribution of bright points. Lawrence \& Cadavid examine phase differences at the part of their FOV where the canopy is located below 400\,km, whereas we examine the areas where the canopy is located below 1600\,km. Given the fact that the \caiih\ line forms below the temperature minimum, within our canopy region there is increased contribution from the undisturbed IN. In fact, examining the phase differences at different canopy heights, the outcome is a distribution similar to those in Fig.~\ref{ca_gb}, but with decreasing inclination and, therefore, our findings are in line with previous results.

\begin{figure*}[ht]
\centering
\includegraphics[width=19 cm]{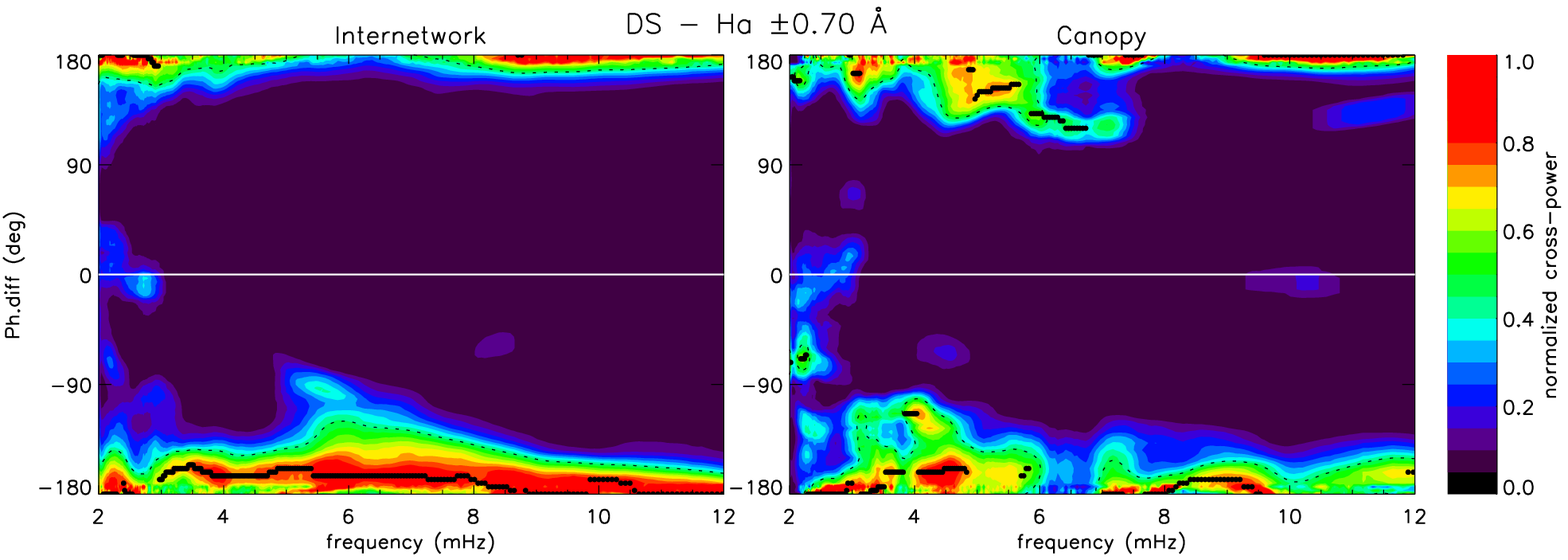}
\includegraphics[width=19 cm]{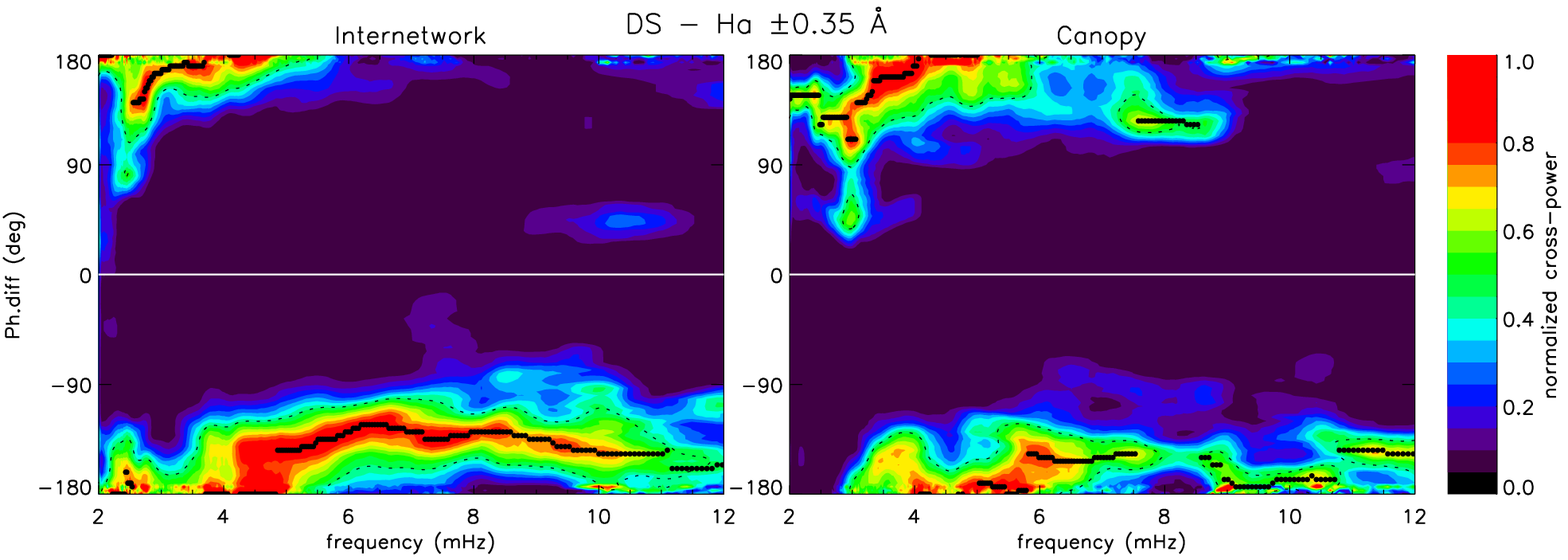}
\caption[]{Same as Fig.~\ref{ds1_ds2}, but for \ha DS and average
wing intensity at $\pm$0.70\,{\AA} (top) and \ha DS and average wing
intensity at $\pm$0.35\,{\AA} (bottom).} \label{ha_vi}
\end{figure*}

\subsection{Phase differences between \ha Doppler Signals}

According to \citet{leenaarts}, the \ha wing at $\pm$0.70\,{\AA} is of photospheric origin (200-600\,km), while the $\pm$0.35\,{\AA} is mostly chromospheric (800-1600\,km), and also contains a significant photospheric contribution. This makes the \ha a valuable, although rather complicated, spectral line to study diverse atmospheric layers. Apart from the fact that there is photospheric contribution in both wing positions, their different behaviour is evident by the different pattern in the 2D maps shown in Fig.~\ref{data}: the DS at $\pm$0.70\,{\AA} at the IN shows impulsive brightenings (upwards motions) and darkenings over a largely uniform background. The $\pm$0.35\,{\AA} DS, on the other hand, shows a more complicated pattern of upwards and downwards motions, of chromospheric origin. Concerning the canopy area, this different behaviour is justified by the power maps in 3 and 5\,min (5.5 and 3.3\,mHz, respectively) presented in \cite{kont10a,kont10b}, where a power halo around the network boundary at the $\pm0.70\,{\AA}$ DS gives its place to a magnetic shadow at the $\pm0.35\,{\AA}$ DS. Having concluded that the two DSs show two distinct regions, that is, below and above the magnetic canopy \citep{kontogiannis14}, we examine wave propagation at this region in comparison with the IN (Fig.~\ref{ds1_ds2} right and left panels, respectively).

At the IN, throughout the entire frequency range, phase differences are positive, meaning that the deeper formed DS leads the higher formed one. Overall, the distribution does not represent dispersion relation typical of wave propagation (monotonically increasing with frequency). This is a common finding when comparison between chromospheric lines is made \citep{mein76,schmieder,fleck89}. Similar to these studies, small positive phase differences are measured. The distribution is thicker at 5.5\,mHz, where crosspower extends to higher positive phase differences, as a possible trail of acoustic wave propagation up to chromospheric layers. According to \citet{fleck_schmitz91}, the chromosphere undergoes resonant excitation at 3\,min as a result of the longer period p-mode spectrum. Higher frequency oscillations, on the other hand, vanish at the chromosphere owing to the process of shock overtaking \citep{fleck_schmitz93}, and transfer their energy to the 3\,min range. This is a purely hydrodynamic process that could take place at the IN. Furthermore, the higher the frequency, the easier for the acoustic waves to undergo conversion to fast magneto-acoustic waves and reflect at low-lying canopies, which are also present at the IN \citep{nutto}. This process might explain detection of negative phase differences above 5\,mHz (as a result of downwards propagating waves) and the decline of the high-frequency phase differences towards zero, which has also been seen in TRACE observations at UV continua \citep{krijger}.

At the canopy (Fig.~\ref{ds1_ds2}, right panel), the behaviour of phase differences is
different. Negative values are measured up to 3.5\,mHz, which means that on long timescales, the \ha$\pm$0.35\,{\AA} DS leads the \ha$\pm$0.70\,{\AA} DS. It is possible that there is an impact of the stochastic appearance-disappearance of mottles on the low-frequency behaviour. In our previous studies, we raised concerns about attributing the 7\,min (2.4\,mHz) period to representing acoustic oscillations. In \citet{kontogiannis14}, we showed that the 7\,min power at the chromosphere shows a different behaviour than that at the 5\,min, and it is very possible that it is not entirely due to magneto-acoustic waves. Conversely, negative phase differences at the low-frequency range may also be attributed to gravity waves \citep{mihalas81}, which have been detected at the upper photosphere \citep{severino03,rutt03}. If these gravity waves exist, they may very easily be affected by the magnetic field of the canopy and convert to either downwards slow magneto-acoustic waves or Alfv\`{e}n waves \citep{newington2010}.

In the 3-4\,mHz frequency range there is a steep increase of phase differences, and the distribution is thicker, with crosspower extending up to higher positive values. This may be explained on the grounds of a significant reduction of the cut-off frequency as a result of waves propagating along inclined magnetic field lines. These low-frequency magneto-acoustic waves are able to propagate upwards in the low-$\beta$ regime, as slow waves through ``magneto-acoustic portals'', which are created by strong and
significantly inclined magnetic fields \citep{jefferies06}. Above the cut-off frequency, phase differences are lower than at the IN and phase differences become equal to 0$\degr$ up to 8\,mHz. This is in total agreement with the results of \citet{nutto}, which indicate that the fast magneto-acoustic waves undergo reflection at the canopy and propagate almost horizontally, causing the adjacent atmospheric layers to oscillate in phase. There is also significant crosspower at negative phase differences, which is consistent with the phase difference maps presented in \citet{kont10a} and have been attributed to downwards propagating waves, after reflection on the magnetic canopy.

\begin{figure*}[ht]
\centering
\includegraphics[width=19 cm]{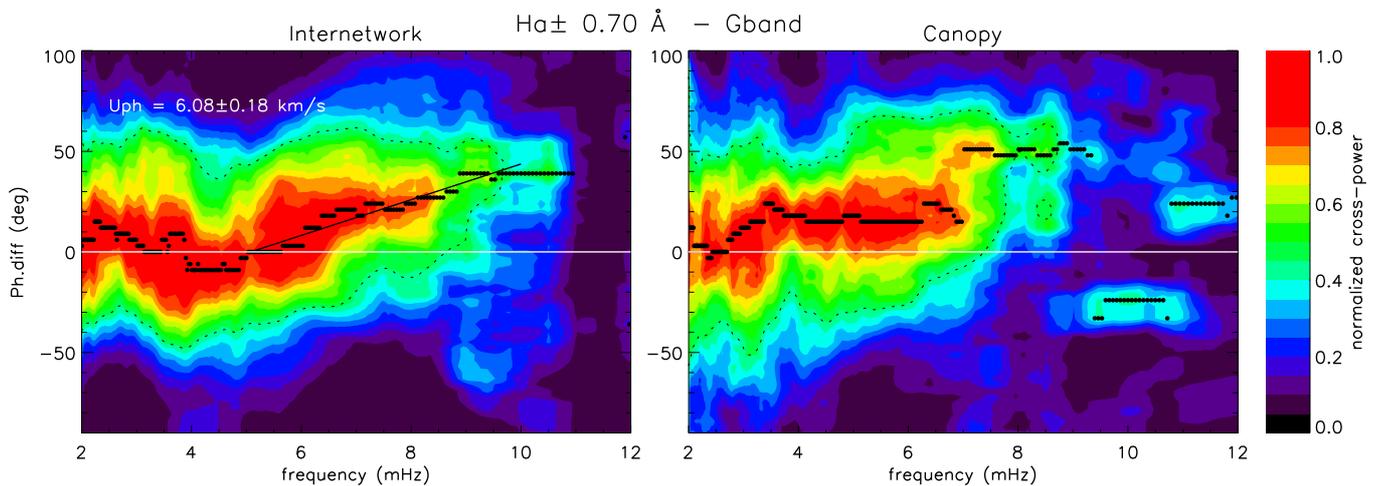}
\caption[]{Same as Fig.~\ref{ca_gb}, but for \Ha$\pm$0.70\,{\AA} --
G-band (top). A height separation of 150\,km is assumed.} \label{ha_gb}
\end{figure*}

\subsection{Phase differences between \ha Doppler Signal and intensity pairs}

Phase differences between intensity (as a proxy of temperature) and velocity may be used to infer the properties of acoustic waves. It is known from the literature that in running waves, temperature and velocity should be largely in phase, while for standing waves, intensity leads velocity by $90\degr$ (adiabatic limit). When heat losses are rapid enough and also in the isothermal case, this phase difference may reach up to $180\degr$ \citep{Whitney58,holweger75}. Negative phase differences between -180$\degr$ and -90$\degr$ indicate downwards propagation.

We use these guidelines to interpret phase differences between the \ha average wing intensities and DS at $\pm$0.35\,{\AA} and $\pm$0.70\,{\AA;} shown in Fig.~\ref{ha_vi}. The DS is not the actual velocity, but a parametric representation of velocity, calculated by intensity values. Regarding phase differences between average wing intensities and DS, crosstalk could be introduced, as shown by \citet{moretti02}. These authors calculate for their data an offset up to 10$\degr$. Furthermore, \ha line formation processes complicate the interpretation of phase differences, since the dependence of the intensity on temperature, established at the outer wings of the line, weakens towards the line centre \citep{leenaarts12}. Finally, regarding ground-based observations, it has been reported that seeing may produce phase differences close to $0\degr$ or $180\degr$ at the high-frequency end of the distribution with increased coherence \citep{endler83}. To see whether there is an influence from seeing, we examined the rest of the crosspower distributions more carefully, before normalization, and found that that there is little or no crosspower concentrated around 180$\degr$. Therefore, we conclude that the signal in the phase differences between the \ha DS and average wing intensity is not due to seeing. In fact, as already mentioned, comparisons of phase differences between \caiih\ and G-band taken from the ground with results based on (seeing-free) observations from space \citep{lawrence12} show remarkable agreement.

Keeping in mind these caveats, how can the measured phase differences be explained? In Fig.~\ref{ha_vi}, positive phase difference means that the intensity leads upwards velocity and vice versa. The distributions at all occasions are concentrated around the $\pm$180$\degr$ line, indicating that velocity and intensity are largely in anti-phase, but there is also significant crosspower at the positive and negative phase difference domains. Phase differences at low and higher frequencies, reaching down to 90$\degr$ may correspond to the evanescent low-frequency waves and also the standing wave pattern of fast waves found around the canopy \citep{nutto}. On the other hand, negative phase differences, between -90$\degr$ and -180$\degr$, correspond to downwards propagating waves. This could be the case for at least part of the crosspower detected in Fig.~\ref{ha_vi}, since the process of conversion and transmission predicts the existence of fast waves directed towards the photosphere. We cannot exclude this possibility, even more since this behaviour is found at frequencies higher than 5\,mHz.

Can we attribute the smaller differences found at the phase differences in the canopy regions to the interaction of the acoustic waves with the magnetic field? \citet{kostik13}, examining a facular area, attribute the scatter of the phase differences throughout the entire range between -$180\degr$ - $180\degr$ to the presence of the magnetic field, while \citet{lites82} also observe lower phase differences in V-I pairs at the network than at the IN (which appears to be the case in the right panels of Fig.~\ref{ha_vi}). Overall, the phase difference distributions resemble those of \citet{kulaczewski} and \citet{deub96} who used chromospheric lines, such as Na\,D1. Deubner et al. state that distributions, such as those observed in Fig.~\ref{ha_vi}, may be explained on the grounds of the superposition of running and standing waves. Having measured the height of reflection of the fast waves and explained the observed oscillatory power on the grounds of mode conversion/transmission \citep{kontogiannis14}, we believe that the distributions in Fig.~\ref{ha_vi} support this scenario. However, as mentioned above, these results should be treated with caution.

\subsection{Phase differences between \ha and G-band.}

Next, we examine phase differences between the G-band and the average \ha wing intensity at $\pm$0.70\,{\AA} from the line centre (Fig.~\ref{ha_gb}). The G-band shows details of the granulation, which is also dominant in the \ha outer wings along with strongly Doppler-shifted absorption features. This fact complicates the comparison with this intensity, introducing noise and reducing coherence, however, a coherent signal shows signs of wave propagation.

At the IN (Fig.~\ref{ha_gb}, left panel), phase differences are concentrated around 0$\degr$ providing evidence of evanescent waves up to 6\,mHz. Then phase differences increase, up to 10\,mHz. Assuming a minimum height separation of 150\,km, the phase speed at the IN is 6.08$\pm$0.18\,km/s, a reasonable value for the acoustic speed at the photosphere and consistent with the corresponding phase speed presented in Fig.~\ref{ca_gb}. At the canopy, phase differences already start to increase  at 3\,mHz, showing the effect of p-mode leakage. Then, there is a plateau up to 7\,mHz, above which the crosspower decreases significantly, although some higher positive and negative phase differences are measured. No regression line was plotted since there is no monotonic increase. From the shape of the distribution, it appears that the interaction with the magnetic canopy affects the phase differences in a manner similar to the \caiih\ -- G-band pair, but more dramatically owing to the fact that the \ha$\pm$0.70\,{\AA} forms below the canopy \citep{kont10b}.

\begin{figure*}[ht]
\centering
\includegraphics[width=19 cm]{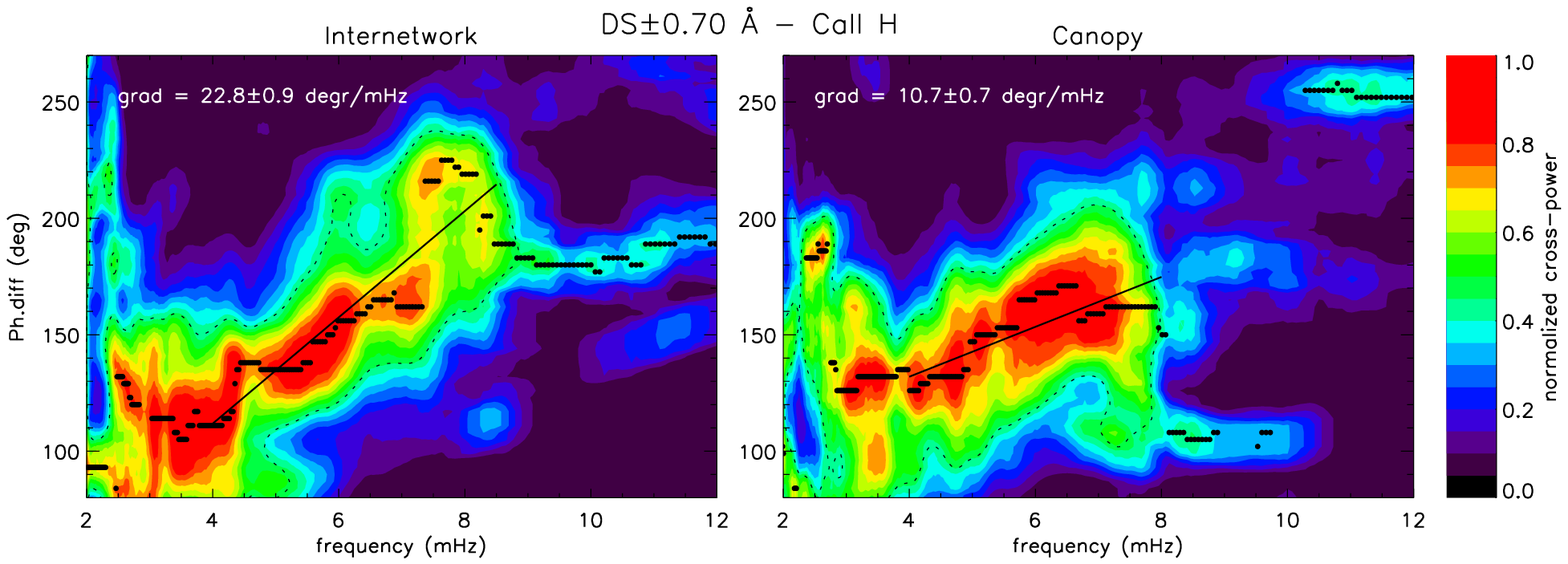}
\includegraphics[width=19 cm]{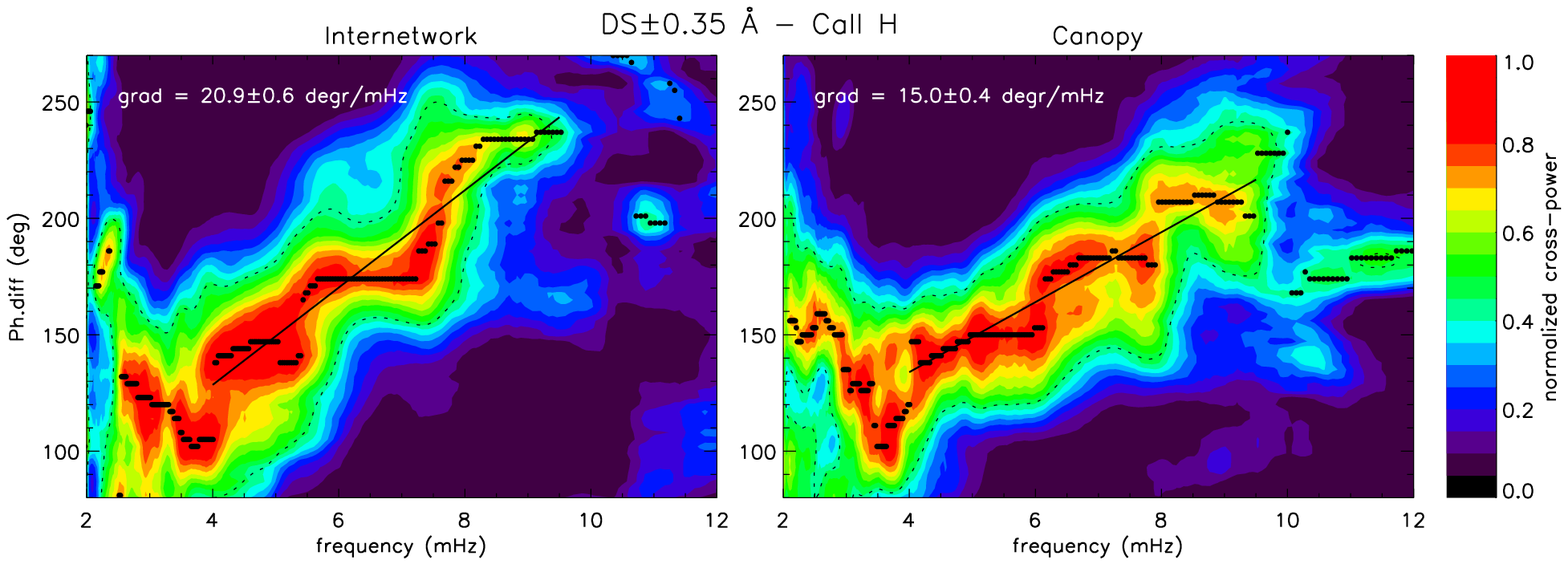}
\caption[]{Same as Fig.~\ref{ca_gb}, for \ha DS at $\pm$0.70\,{\AA}
and \caiih\ intensity (top row) and \ha DS at $\pm$0.35\,{\AA} and
\caiih\ intensity (bottom row).} \label{ca_ds}
\end{figure*}

\subsection{Phase differences between \Ha Doppler shifts and \caiih\ intensities.}

In phase difference analysis, velocity and intensity pairs from the same spectral line are usually compared, although in some cases, a comparison of velocities and intensities from different spectral regions has been used \citep{cram,judge}. Motivation to perform a comparison between the \caiih\ intensity and  \ha DS came from inspecting the \ha DS time series along with the \caiih\ intensity images. Localized, short duration, upwards motions are detected abundantly at the IN in \ha DS time series, which are reminiscent of the pattern of the intensity variations in \caiih, superposed on the reversed granulation pattern. According to \citet{rutten08}, the \ha wings show a response to the IN acoustic shocks that are abundant in the \caiih. To check the response of the \ha DS to the acoustic waves found at the upper photosphere, we constructed the halftone images of Fig.~\ref{ca_ds}.

In all panels, phase differences exhibit a monotonic increase between 4-8\,mHz starting above 100$\degr$--120$\degr$. Through observations and theory \citep{lites93,skart94} it has been found that the \caiih\ intensity leads the velocity in the same line by 100$\degr$--120$\degr$ at the acoustic range. Based on this result, one might assume that a comparison between the \caiih\ velocity and the \ha DS would give a similar distribution starting at 0$\degr$. Therefore, we conclude that there is acoustic wave propagation at the IN between the levels where \caiih\ and \ha DS are formed. Given the fact that the IN is dominated by acoustic shocks, as suggested by \citet{cs97}, this provides an observational evidence of the \ha velocity response to these IN acoustic shocks.

This wave propagation (monotonic increase in phase difference) commences at frequencies lower than the typical acoustic cut-off. Furthermore, the gradient of the distribution of the $\pm$0.70\,{\AA} DS --\caiih\ is slightly higher than the $\pm$0.35\,{\AA} DS--\caiih\ pair, which is consistent with a larger height separation in the former case than in the latter. However, the similar values of the gradients demonstrate that this difference in height separation does not contribute dramatically to the measured phase differences, which is in agreement with the small phase differences presented in Fig.~\ref{ds1_ds2} between the two DS of H$\alpha$. At the magnetic canopy, these gradients are even lower, which is also consistent with the reflection of the fast magneto-acoustic waves near the network boundaries.

A comparison between high temporal, spatial, and spectral resolution observations in \ha and \caiih, which are becoming increasingly available to the community, will lead to valuable results on the connection
between the acoustically dominated upper photosphere/lower chromosphere at the temperature minimum region and the \ha line formation at the IN.

\section{Conclusions and discussion}

We have examined phase differences of oscillations using time series of co-temporal photospheric and chromospheric images of a solar quiet region. We conclude that the small-scale magnetic field of the quiet Sun affects the phase differences of oscillations in the various heights sampled by our data, in the context of magnetoacoustic mode conversion/transmission. This effect was studied in the past as
``network-IN distinctions'', but it now seems that the crucial distinction should be ``canopy-IN''. As already mentioned, the traditional view of the IN as a field-free component of the solar atmosphere is itself under revision in light of the latest high-resolution observations of the solar magnetism \citep{lites08}.

In most of the cases examined in this study, the acoustic cut-off frequency is lower than that expected for a gravitationally stratified atmosphere that is free of magnetic
fields. Figures~\ref{ca_gb},~\ref{ds1_ds2}, and \ref{ha_gb} show evidence of wave propagation for frequencies as low as 3\,mHz, at the canopy and in some cases at the IN, with a phase speed roughly equal to the local sound speed. The importance of this mechanism, as illustrated also in numerical studies, is the fact that slow waves are easily channelled to the upper atmospheric layers, travelling along the magnetic field lines.

Above the acoustic cut-off frequency, wave propagation is detected between G-band - \caiih, G-band - H$\alpha$ and \ha DS - \caiih. The presence of the magnetic canopy appears to alter wave propagations by systematically lowering the gradient of the phase difference distributions, hence, increasing the phase speed. In our context, this is an indication of wave energy being transferred to the fast magneto-acoustic mode, combined with the superposition of a standing wave pattern on the existing acoustic wave field. This idea has been speculated in the past by \citet{deub_fleck90}, but their data analysis lacked simultaneous high-resolution magnetograms. \citet{kontogiannis14}, based on the dependence of the oscillatory power on the magnetic field inclination, demonstrated that mode conversion and transmission takes place at the magnetic canopy. As already mentioned, conversion to fast magnetoacoustic waves is more efficient for higher frequencies and produces shorter phase differences due to their higher phase speed (fast speed). This may explain the flattening of the crosspower distribution in higher frequencies, at the magnetic canopy (Fig.~\ref{ha_gb}), while the presence of small-scale and unresolved magnetic fields may be invoked to explain this effect, even at the undisturbed IN where acoustic wave propagation does not seem to persist for high frequencies. In the context of the proposed mechanism, the magnetic field acts as a directional filter, which eventually allows only acoustic disturbances travelling along the magnetic field lines. As for the energy carried by the fast waves, in the 3-D case it has been shown that Alfv\'{e}n waves are produced at the sites of reflection \citep{khom12}. We believe that our measurements of phase differences between intensity and DS in \ha support this scenario, but this specific subject should be revisited using several high-resolution, ground-based observations of the chromosphere.

Our analysis also reveals that the \ha DS exhibits a response to the phenomena occurring in the temperature minimum region, which is sampled by the \caiih\ line. \citet{leenaarts12} have shown that the \ha line is decoupled from temperature, above 1\,Mm and is sensitive to density variations. Acoustic shocks produce such enhancements in density \citep{cs97} at the IN and mostly below the magnetic canopy
and lead to DS variations in H$\alpha$. These acoustic shocks have been found to avoid magnetic field concentrations at the IN and the network \citep{vecchio09}, and magnetic shadows are an example of this phenomenon.
The effect of mode conversion and transmission is also visible here, as discussed in the previous section.

When interpreting results of phase difference analysis, difficulties arise from the width in the contribution functions of the bandpasses used, the vagueness of the HOF relating to the
corresponding physical and magnetic conditions, and the uncertainty/variety of the solar conditions across the FOV. In this study, we assumed some average values for the height separation of the layers sampled by our data based on the estimated formation heights of \ha \citep{leenaarts} and \caiih\ \citep{rutten2004b}. Non-LTE effects involved in the formation of \ha along with the limited spectral coverage of the \ha profile in our data does not allow us to completely disentangle the contribution of the line width,  opacity, and line shift. Therefore, results based on the \ha line should be treated with caution, particularly those concerning the interpretation of V-I phase spectra. These spectra are even more complicated taking into account that the determination of the average wing intensities and DS may introduce crosstalk between velocity and intensity signals \citep{moretti02}. This is something worth revisiting in the future in light of improved spectral and temporal resolution chromospheric observations.

To our defence and as far as the validity of our approach is concerned, the confirmation of previous results concerning the \caiih-G-band intensity pair \citep{rutten2004b,lawrence12} provides assurance that the determination of phase differences itself, through our approach, is accurate. There are, however, limitations in our method: the oblique propagation of slow and fast waves (along and perpendicular to the magnetic field) imposes a selection effect in favour of vertically propagating disturbances. In order to fully identify the relation between oscillations at the lower and upper parts of the chromosphere, one must be able to combine large statistical significance \citep[e.g. as in][]{dewijn09} with a localized treatment that may highlight the effect of the diversity of the magnetic field configuration, taking  the magnetic field inclination into
account, for example  \citep[in a manner similar to the work of][]{bloomfield07}. Admittedly, inside the canopy area, the inclination varies, since the magnetic field is largely vertical near the network and progressively more inclined towards the outer parts of the canopy. Since the transmission/conversion of magneto-acoustic waves depends on the inclination of the magnetic field, an examination of the phase differences variation in regions of different inclination angles would be ideal. However, it is difficult to achieve a statistically significant result in the fine magnetic concentrations of the network, as opposed to the extended active regions. Data sets of several network areas would be required, including simultaneous high-resolution magnetograms to amass a large statistical sample. Our analysis demonstrates the overall effect of the magnetic canopy on the wave propagation, but we believe that analysis of higher spectral and spatial resolution data will shed more light on the propagation of waves through the magnetized chromospheric plasma.

\begin{acknowledgements}
The observations have been funded by the Optical Infrared
Coordination network (OPTICON, http://www.ing.iac.es/opticon), a
major international collaboration supported by the Research
Infrastructures Program of the European Commission's sixth
Framework Program. The research was funded through the project
``SOLAR-4068'', which is implemented under the ``ARISTEIA II''
Action of the  operational program ``Education and Lifelong
Learning'' and is cofunded by the European Social Fund (ESF) and
Greek national funds. The DOT was operated at the Spanish
Observatorio del Roque de los Muchachos of the Instituto de
Astrof\'{i}sica de Canarias. The authors thank P. S\"{u}tterlin for
the DOT observations and R. Rutten for the data reduction. Hinode is
a Japanese mission developed and launched by ISAS/JAXA,
collaborating with NAOJ as a domestic partner, and NASA and STFC
(UK) as international partners. Scientific operation of the Hinode
mission is conducted by the Hinode science team organized at
ISAS/JAXA. This team mainly consists of scientists from institutes
in the partner countries. Support for the post-launch operation is
provided by JAXA and NAOJ (Japan), STFC (U.K.), NASA, ESA, and NSC
(Norway). Hinode SOT/SP Inversions were conducted at NCAR under the
framework of the Community Spectro-polarimetric Analysis Center
(CSAC; http://www.csac.hao.ucar.edu).

\end{acknowledgements}

\end{document}